\documentclass[aps,prl,twocolumn,showpacs,letterpaper,superscriptaddress]{revtex4-1}

\usepackage{graphicx,dcolumn,longtable,epsfig}
\usepackage[usenames]{color}
\usepackage{amssymb}
\usepackage{amsmath}
\usepackage{epstopdf}
\usepackage{bm}

\input epsf.tex
\newcommand{\etal}{{\it et al.}}

\def\pra#1#2#3{Phys.~Rev.~A~{\bf #1},\ #2\ (#3)}
\def\prb#1#2#3{Phys.~Rev.~B~{\bf #1},\ #2\ (#3)}

\def\prl#1#2#3{Phys.~Rev.~Lett.~{\bf #1},\ #2\ (#3)}

\def\nat#1#2#3{Nature~{\bf #1},\ #2\ (#3)}

\def\natp#1#2#3{Nat.~Phys.~{\bf #1},\ #2\ (#3)}

\def\etal{{\it et al.}}
\def\bea{\begin{eqnarray}}
\def\eea{\end{eqnarray}}
\def\be{\begin{equation}}
\def\ee{\end{equation}}

\usepackage{color}

\begin{document}
\title{First order phase transitions in optical lattices with tunable three-body onsite interaction}
\author{ A.~Safavi-Naini}
\affiliation{ITAMP, Harvard-Smithsonian Center for Astrophysics, Cambridge, Massachusetts, 02138, USA}
\affiliation{Department of Physics, Massachusetts Institute of Technology, Cambridge, Massachusetts, 02139, USA}
\author{  J.~von  Stecher\footnote{Current address: Tech-X Corporation, Boulder, Colorado 80303, USA}}
\affiliation{JILA, National Institute of Standards and Technology and University of Colorado, Boulder, Colorado, 80309, USA}
\author{ B.~Capogrosso-Sansone }
\affiliation{ITAMP, Harvard-Smithsonian Center for Astrophysics, Cambridge, Massachusetts, 02138, USA}
\affiliation{Homer L. Dodge Department of Physics and Astronomy, The University of Oklahoma, Norman, Oklahoma ,73019, USA }
\author{ Seth T.~Rittenhouse }
\affiliation{ITAMP, Harvard-Smithsonian Center for Astrophysics, Cambridge, Massachusetts, 02138, USA}

\begin{abstract}
We study the two-dimensional Bose-Hubbard model in the presence of a three-body interaction term, both at a mean field level and via quantum Monte Carlo simulations.
The three-body term is tuned by coupling the triply occupied states to a trapped universal trimer.
 We find that,
 for sufficiently attractive three-body interaction the $n=2$ Mott lobe disappears and the system displays first order phase
transitions separating the $n=1$ from the $n=3$ lobes, and the $n=1$ and $n=3$ Mott insulator from the superfluid.
We have also analyzed the effect of finite temperature and found that transitions are still of first order at temperatures $T\sim J$ where $J$ is the hopping matrix element.
\end{abstract}

\date{\today}
\maketitle

The Bose Hubbard (BH) model and its second order superfluid (SF)-Mott-insulator (MI) transition represent one of the paradigmatic examples of strongly interacting many-body physics in lattice structures~\cite{Fisher}. The unprecedented control over ultra-cold atoms in optical lattices not only allows for a clean experimental realization of the BH model~\cite{Zoller,Bloch_RMP}, but also the exploration of a panoply of quantum effects beyond the standard BH model (see references \cite{DeMarco,Schneble,Ospelkaus,Esslinger2,disorder1,disorder2,mixtures1,mixtures2,tomasz}).

One key element for such impressive progress is the possibility of tuning two-body interactions by using Fesh\-bach resonances or changing the strength of the lattice confinement. More recently, effective multi-body interactions have been experimentally observed~\cite{Bloch2010,Nagerl2011}. The question that naturally arises is how these interactions affect the many-body behavior. Topological phases such as fractional quantum Hall states appear as ground states to model Hamiltonians with strong three-body interactions while exotic quantum phases have been predicted for bosonic Hamiltonians with many-body interactions, such as the ring exchange model \cite{FQH,ringex}. An important first step in realizing these models using ultra-cold atoms was the recognition that strong three-body losses lead to an effective hard core three-body interaction that can be used, for instance, to stabilize the BH model with attractive two-body interactions~\cite{Daley2009}. Under these conditions, the system can undergo a {\it first} order MI to SF transition in the presence of strong pairing interactions~\cite{Kuklov}. Despite these recent studies, lattice systems with three-body interactions remain largely unexplored.

\begin{figure}
\begin{center}
\includegraphics[width=0.5\textwidth]{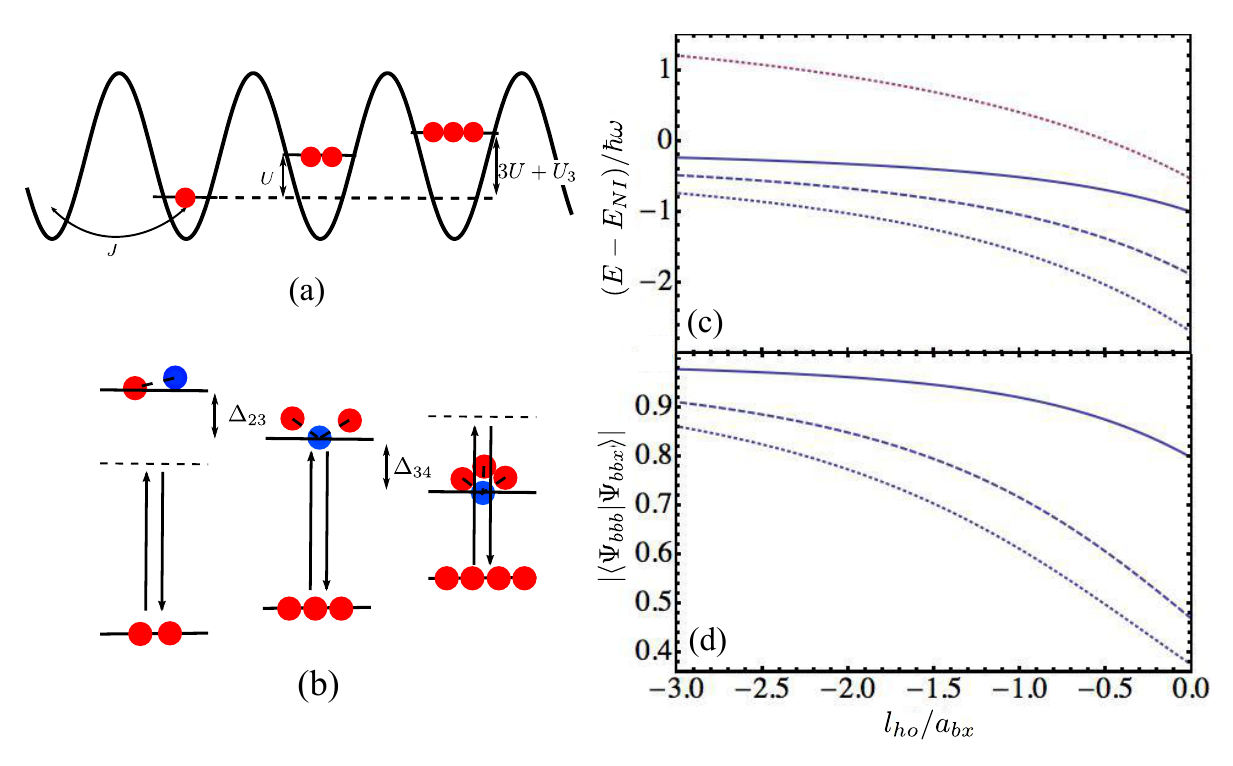}
\caption{Schematic  representations of (a) the Bose-Hubbard Hamiltonian considered here [Eq.~\eqref{eq:BH3}] and (b) the rf field tuned three-body onsite interaction are shown. In (b) red and blue circles represent bosons in the lowest and (excited) hyperfine states $(b$ and $x$) respectively.  The dotted line represents the energy the rf field is tuned to with respect to the excited two- and four-body states.  (c) The energy of the lowest two-, three- and four-body states (solid, dashed and dotted lines respectively) in the excited hyperfine state is shown in trap units with respect to the non-interacting energy $E_{NI}$.  Also shown is the energy of the first excited four-body state (upper dotted line) to demonstrate that the universal three-body state is in fact isolated. (d) The wavefunction overlap between the non-interacting ground state and the lowest excited hyperfine state is shown as a function of $l_{ho}/a_{bx}$ for two, three and four atoms (solid, dashed and dotted lines respectively). In both (c) and (d) the calculations are done for model Rb atoms described in the text.}\label{fig:schem}
\end{center}
\end{figure}

In this Letter, we analyze how the many-body physics of the BH model is affected by the presence of local and tunable three-body interactions.
First, we propose a mechanism for engineering a three-body onsite interaction term, $U_3$, which is controlled by an external rf pulse that couples the triply occupied state with a three-body bound state associated with an excited hyperfine state. This local three-body interaction only affects triply occupied sites leading to a modified BH Hamiltonian
\begin{equation}\label{eq:BH3}
H=-J\sum_{\langle i, j \rangle} a_i^{\dagger}a_j +\sum_i [\frac{U}{2} n_i(n_i-1)+\delta_{n_i,3}U_3-\mu n_i]
\end{equation}
where $a^{\dag}_i(a_i)$ is the bosonic creation (annihilation) operator, $n_i = a^{\dag}_i a_i$, $J$ is the hopping matrix element, $U$ is the two-body onsite interaction, $U_3$ is the $3$-body onsite interaction, $\mu$ is the chemical potential, and ${\langle i, j \rangle}$  denote sum over nearest neighbor sites only [see Fig.~\ref{fig:schem}(a)].
Note that the three-body interaction considered here is different from the more conventional interaction of the form $U_3 n (n-1)(n-2)/6$.

Next, we use a mean-field Gutzwiller approach to study the BH Hamiltonian in the presence of such three-body interaction in the $U>0$, $U_3<0$ regime. We focus on the $|U_3|>U$ region where the $n=2$ ($n=4$) lobe disappears. In this regime a direct first order phase transition at finite hopping can occur between the $n=1$ and $n=3$ lobes. Quantum Monte Carlo (QMC) simulations~\cite{Worm} confirm the existence of a first order phase transition and provide quantitative predictions of the phase diagram
in two dimensions and particular values of $U$ and $U_3$. Finally, we briefly discussed finite temperature effects and experimental signature of the first order transition.

To achieve a separately tunable, on-site, three-body interaction of the form shown in Eq.~\ref{eq:BH3}, we envision a system in which a universal three-body bound state is attached to an excited hyperfine threshold  \cite{BraatenHammer} which is coupled to the identical boson ground state by an external rf field (shown schematically in Fig.~\ref{fig:schem}(b))  \cite{RFexp,RFtheory}. In this scheme, identical bosons in two different hyperfine states (labeled $b$ and $x$ for the lowest state and an excited state respectively) sit on a single site which we model as an isotropic oscillator with oscillator frequency $\omega$ length $l_{ho}$. 
For a three-body bound state to form in an excited hyperfine state, we consider a system with repulsive $bb$ non-resonant interaction ($0<a_{bb}\ll l_{ho}$)  and large $bx$ scattering length ($a_{bb}\gg |a_{bx}|$).  In this situation, universal three-body Efimov states form attached to the $bbx$ three-body hyperfine state~\cite{BraatenHammer}.

To analyze this scenario, we explore  two-, three- and four-body single site physics within the harmonic approximation using a model short range interactions and a correlated Gaussian basis set expansion~\cite{suzuki1998sva,Droplets}. We tune the interaction parameters to achieve a $bbx$ Efimov trimer whose binding energy ($E_b=E_{NI}-E$ where $E_{NI}$ is the non-interacting energy of the trapped system) is comparable to the trapping energy $\hbar\omega$ (see Fig.~\ref{fig:schem}(c)). This particular scenario is suitable for achieving the proposed Hamiltonian for two reasons. First, the two-, three- and four-body binding energies are well separated allowing to tune the rf pulse in resonance with a particular  single site occupancy; and second, the large wavefunction overlaps [see Fig.~\ref{fig:schem}(d)] imply an efficient rf transition probability.
Our numerical calculations show that the lowest three-body energy in the $bbx$ configuration is lower than the two-body energy ($bx$ configuration), and that
%
   for each Efimov trimer state, there is a single four-body state ($bbbx$ configuration) bound below the trimer state.  All other four-body states lie above the trimer energy. This is in contrast to the more commonly considered case of four identical bosons in free space in which there are both a deeply bound and a weakly bound tetramer associated with each Efimov state \cite{4bodyEfimov}.  The energies for two-, three-, and four-body states are shown schematically in Fig.~\ref{fig:schem}(b) along with the energies at which the rf field is tuned to (dotted lines).

By detuning the rf field to the red of the Efimov state for three body occupation, both the two- and three-body identical boson ground states are shifted down.  However, because there is an energy difference between the two- and three-body transition energy the two-boson state is shifted significantly less than the three-body.  For higher occupation numbers, the rf-field is far blue detuned from the bound states and thus the four- or more identical boson ground states experience a weak upward shift.  One might expect an additional shift in the higher occupation number states (four or more) resulting from a three-boson and spectator particle like system.  However, for these weakly bound universal Efimov states, the size of the three-body state is similar to the trapping length, and thus additional bosons on site interact with the Efimov state and shift the resulting N-body excited state energy off resonance with the rf field.

As an initial study we consider ${}^{85}$Rb.  Since resonance structure for scattering between hyperfine state is not known, we will assume that there exists an s-wave scattering resonance between the lowest and first excited hyperfine states at some external magnetic field strength. For simplicity, we consider that identical bosons are roughly noninteracting.  Assuming that the energy of the Efimov state is determined by the Van der Waals length of Rb, $r_{vd}\approx 82$ a.u.  \cite{BraatenHammer},  and a lattice site trapping frequency of $\omega = 2 \pi \times 10$ kHz, an Efimov state will arise at $E_{3B}\approx -2 \hbar \omega$ with respect to the $bbx$ non-interacting energy. Under these circumstances, we predict that $U_3$ can be tuned to be attractive and of order $U$ with a detuning of $\Delta \sim 1000$kHz from the Efimov state transition energy.   This large detuning also serves to mitigate the generally short lifetimes of Efimov states(on the order of $~10 \mu$s \cite{lifetimes}).

This initial investigations, presented above as a plausibility argument, indicates that using the above scheme is feasible with existing experimental techniques. 
A more detailed study to determine the effects of rf coupling to an excited three-body state is left for future investigations \cite{future}. Additionally, direct rf association of universal trimer states has already been demonstrated in ultra-cold, three-component, Fermi and Bose gases \cite{RFexp}  lending credibility to the experimental accessibility of this model.

\begin{figure}
\begin{center}
\includegraphics [width=0.45\textwidth]{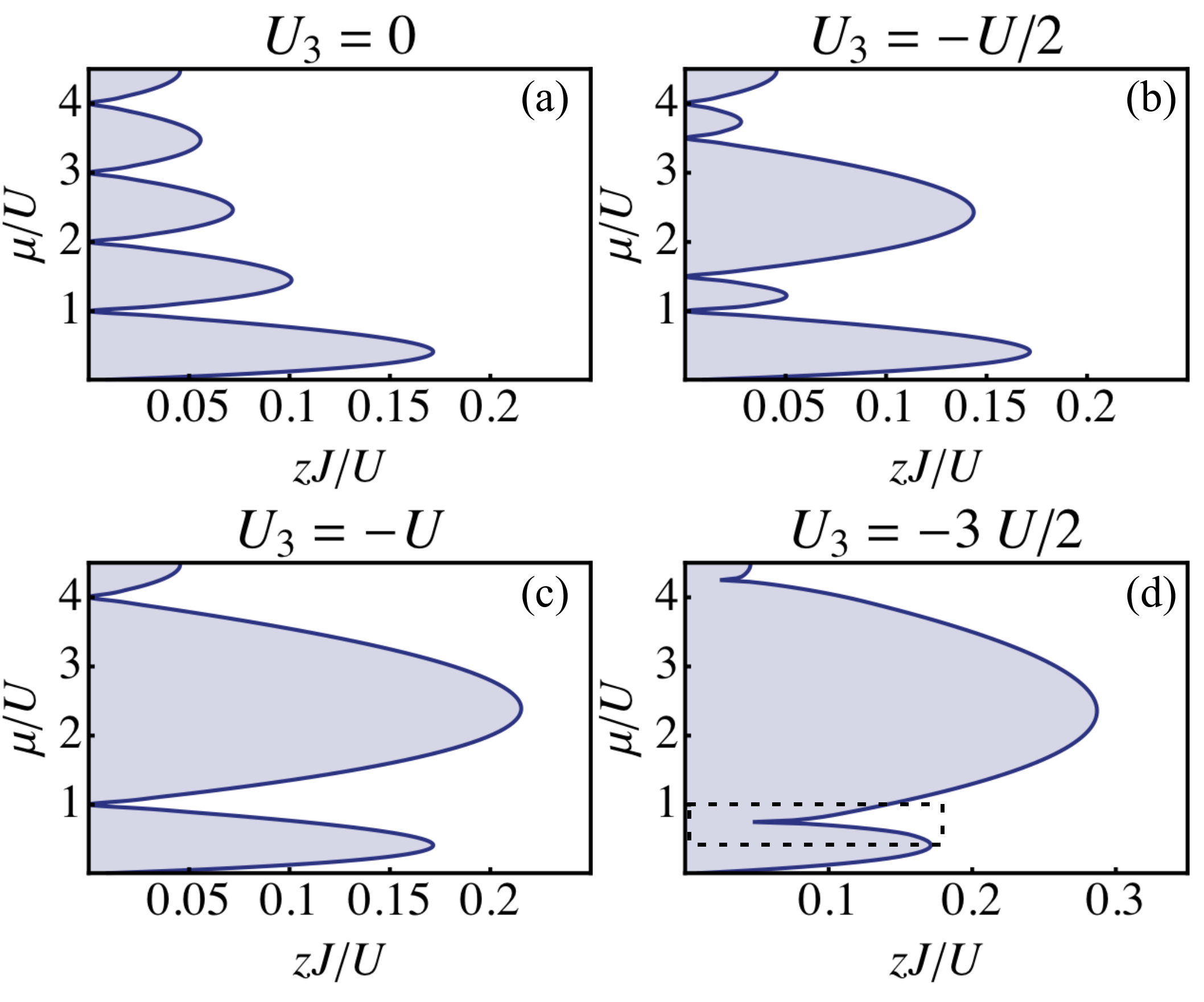}
\caption{
(Color online) Mean field phase diagram in the $\mu/U$ vs. $zJ/U$ plane for $\vert U_3\vert=0$ where $z=4$ is the number of nearest neighbours in two dimensions. (a),
$\vert U_3\vert=U/2$ (b), $\vert U_3\vert=U$ (c), and $\vert U_3\vert=-3U/2$ (d). When $0<\vert U_3\vert<U$ the $n=2$ and $n=4$ lobes visibly shrink in favor of the $n=3$ lobe,
until they completely disappear at $\vert U_3\vert=U$ (c).  For  $\vert U_3\vert>U$ the $n=3$ lobe begins to overlap with the $n=1$ and $n=4$ lobes and a direct phase transition between MI lobes becomes possible. The dotted rectangle in (d) highlights the region examined in detail in Fig.~\ref{fig:MF-QMC}.}\label{fig:var}
\end{center}
\end{figure}
We will now use Gutzwiller mean field theory to study the modified BH Hamiltonian described by Eq.~\eqref{eq:BH3}.\\ \indent
The Gutzwiller mean field theory is constructed by replacing the full Hamiltonian by an effective local Hamiltonian subject to a self-consistency condition. We introduce the superfluid order parameter $\psi=\langle a_i^{\dagger}\rangle=\langle a_i\rangle$ and the Gutzwiller wavefunction $\vert G \rangle= \Pi_{i=0}^{N} ( \sum_{n=0}^{\infty}f_n^{(i)}\vert n_i\rangle)$, so that the effective Hamiltonian for a translationally invariant system, i.e.  $f_n^{(i)}=f_n$, takes the form,
\begin{equation}\label{eq:Gfunctional}
E[\psi]=-Jz\psi\sum_n\sqrt{n+1}(f_{n+1}^{\star}f_n+c.c.)+zJ\psi^2+E_n
\end{equation}
where $z$ is the coordination number, $f_n$ are variational parameters, and $E_n=\frac{U}{2}n(n-1)+\delta_{n,3}U_3-\mu n$.
The problem is now reduced to determining the set of coefficients $\{f_n\}$ which minimize $E[\psi]$ and satisfy the normalization condition $\langle G\vert G \rangle =\sum_n\vert f_n\vert^2=1$~\cite{Zoller,Gutz1,Gutz2}.\\ \indent
Figure 2 shows the ground state phase diagram of
model~(\ref{eq:BH3}) at different values of $U_3$. As we increase the magnitude of $|U_3|$ from 0 to $U$ the $n=2$
and $n=4$ Mott lobes shrink considerably while the $n=3$ lobe increases in size [as seen in Fig.~\ref{fig:var}(b)]. In particular, for $U_3\geqslant-U$
the $n=2$ and $n=4$ lobes completely disappear since it is now energetically more favorable to have occupation
number $n=3$ [Fig.~\ref{fig:var}(c)]. This can be easily understood in the zero hopping limit. At $\mu=\mu_{12}=U$ a doubly occupied site has the same
energy as a singly occupied one. At $\mu=\mu_{13}=(3U+U_3)/2$, instead, a singly occupied site has the same
energy as a triply occupied one. The condition $\mu_{13} \le \mu_{12}$ sets the $U_3$ value for which the second lobe disappears,
i.e. $|U_3| > U$ (at $|U_3|=U$, sites with occupation number $n=1,2,3$ are degenerate in energy for $\mu=\mu_{12}=\mu_{13}$). Direct transitions
from MI occupation numbers $n=1$ to $n=3$ survive at finite hopping [Fig.~\ref{fig:var}(d)] confirmed below using QMC.
The same argument shows that
$|U_3|>U$ also implies the disappearance of the
4th lobe. One can easily see that upon further increasing $U_3$, all lobes other than $n=3$ will eventually
disappear, e.g. at $U_3=-3U$, the $n=1$ and $n=5$ lobes disappear.
\\ \indent
We have monitored the behaviour of mean field energy [Eq.~\eqref{eq:Gfunctional}] at fixed $\mu/U$ while varying $J/U$ to study the order of phase transitions described by model ~\eqref{eq:BH3}.
The formation of double minima structure in the mean field energy functional $E[\psi]$ is a signature of first order phase transitions. We have observed such double minima structure at $U_3=-1.5U$ for the
$n=1$ MI-SF and the $n=3$ MI-SF transitions. The occurrence of first order transitions can be
understood with a simple argument. At fixed small $J/U$, $|U_3|\sim U$, and upon increasing (decreasing) $\mu$ in order to dope the $n=1$ ($n=3$)
MI with particles (holes), double occupancy will be suppressed in favor of triply occupied sites. At large enough $|U_3|$ such mechanism
will eventually prevent a gradual addition (subtraction) of particles resulting in first rather than second order transitions.
Second order transitions will be restored at large enough $J/U$ as the kinetic energy gain due to hopping of extra particles
(holes) will again favor a gradual change in density.

\begin{figure}
\begin{center}
\includegraphics [width=0.45\textwidth]{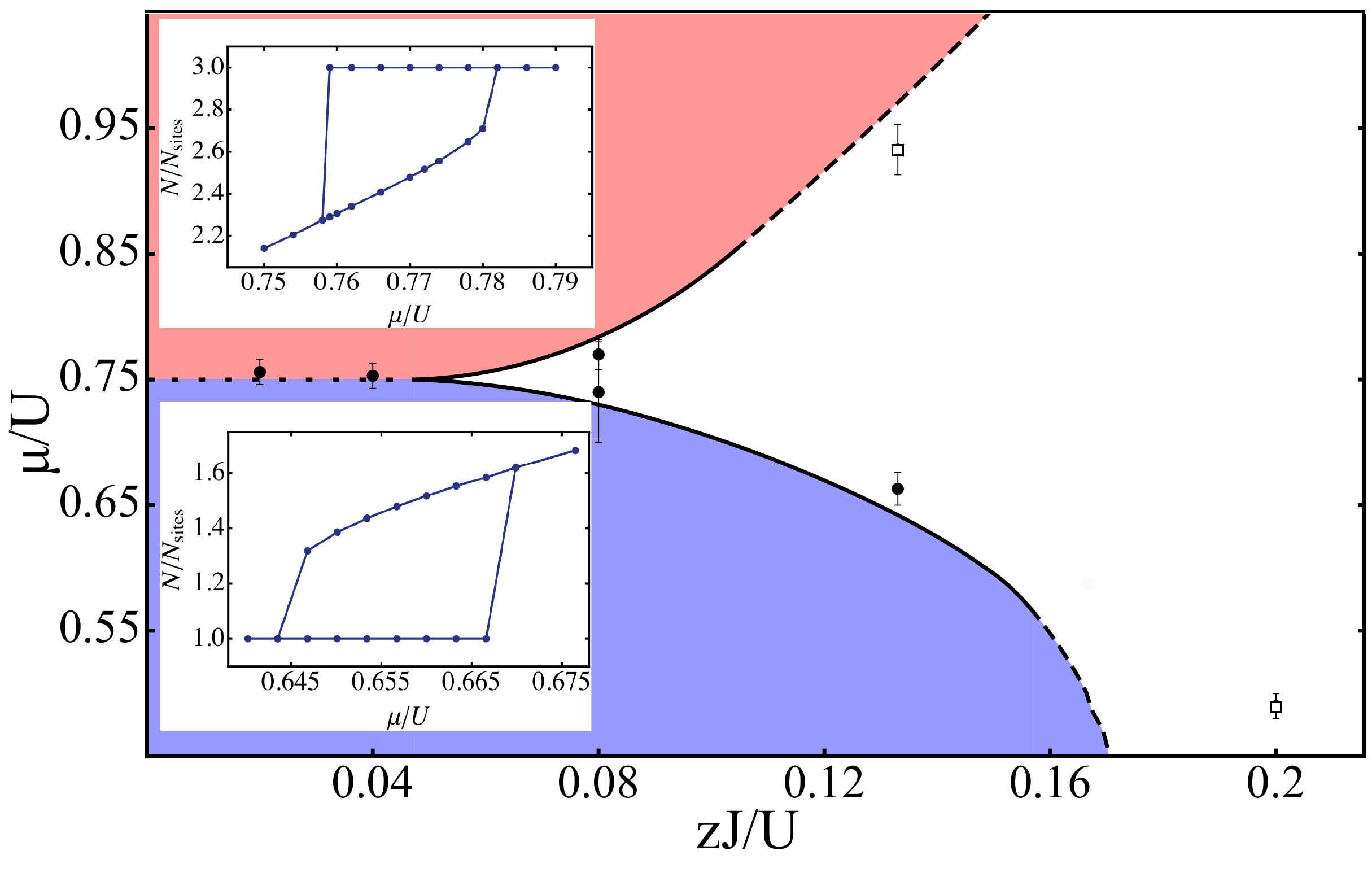}
\caption{First order phase transitions at $U_3=-1.5U$.The  dotted line corresponds to the first order phase transition from the $n=1$ MI (blue) to $n=3$ MI (pink) predicted by mean field theory.
Solid lines refer to mean field first order transitions from MI to SF. Solid circles are QMC results
from hysteretic curves. Dashed lines correspond to second order mean field MI-SF transitions. Open squares are second order transition points from QMC.
Lower (upper) inset shows examples of hysteretic behavior for the $n=1$ ($n=3$) MI-to-SF
transition.
}\label{fig:MF-QMC}
\end{center}
\end{figure}

In order to confirm the mean-field predictions, we have performed QMC simulations on a square
lattice of linear size up to $L=24$ (and $L=30$ in certain cases) for selected values of
$J/U$, and at $\beta=\frac{1}{k_BT}=L/J$ which corresponds to effective zero temperature regime.
Figure ~\ref{fig:MF-QMC} compares the QMC results with the mean-field predictions of the phase diagram for $U_3= -3U/2$.
As mentioned above, a direct transition from $n=1$ MI (lower lobe) to $n=3$ MI (upper lobe) survives at finite hopping.
This first order transition is depicted by the dotted line,
while solid lines refer to first order transitions from MI to SF. The solid and open symbols correspond to QMC predictions of the phase boundary with first and second order phase transitions.

To extract transition points, we have analyzed the particle density $n$ as a function of $\mu$.
Additionally we have performed hysteresis analysis by sweeping back and fourth in chemical potential and calculating
the corresponding particle density.
The hysteretic behavior of the system along the phase boundaries (solid symbols) further
confirms that these are first order transitions.
We show two examples of such curves for the $n=1$ MI-SF
and $n=3$ MI-SF transitions, in the lower and upper inset respectively.
Based on the energy argument previously discussed, we expect
the phase transition to become second order as
$J/U$ is increased. Indeed, larger kinetic energy will favor formation of particle/hole excitations on top of the MI.
The energy gain due to hopping of the latter will compete with the attractive three-body interaction and will eventually
restore the second order MI-SF transition driven by addition/subtraction of small number of particles from the MI regime.
\\ \indent We have used QMC simulations to benchmark the J/U
values at which first order phase transitions become second order.
Second order MI-SF transitions are depicted in Fig.~\ref{fig:MF-QMC}  by dashed mean field lines and open squares representing QMC results. For $z=4$, the
$n=1$ ($n=3$) MI-SF transition becomes of second order at $zJ/U=0.20\pm0.02$ ($zJ/U=0.133\pm0.02$).
We estimate the position triple point using the mean-field approximation [Eq.~\eqref{eq:Gfunctional}] where we truncate the Hilbert space to the $n_i=1,2,3$ states.
Using this approximation we find $\left(\frac{zJ}{U}\right)_{TP}=-(U+U_3)/10$. For $U_3=-1.5U$ and $z=4$ this gives $J/U = 0.05$.
\\ \indent
First order phase transitions present in our model can be experimentally detected due to a loss of adiabaticity across the phase boundary even upon an arbitrarily slow ramping up or down of the optical lattice, as suggested in~\cite{Rey} or by observing the hysteretic behavior. In addition, first order phase transitions are characterized by discontinuity in density profiles, a local observable easily accessible with state of the art techniques~\cite{Chin}.
\\ \indent Finally we have looked at how first order phase transitions are affected by finite temperature.
Strictly speaking, the MI state exists only at zero temperature. In practice, MI features persist up to
temperature $T\sim 0.2U$~\cite{Gerbier}.
QMC results show that phase transitions are still of first order at temperatures of $T\sim J$, where MI features are still well defined.
A more extensive study of the behavior of the system at finite temperature will be the subject of future investigations.
\\ \indent Concluding, we have studied an extended version of the Bose-Hubbard model which includes an attractive three-body interaction term $U_3$, both at a mean field level and by means of quantum Monte Carlo simulations. The three-body term results from  a universal three-body bound state attached to an excited threshold and can be tuned via an external rf field. We have found that at $|U_3|>U$, where the $n=2$ lobe disappears, there exists a first order phase transition separating the $n=1$ from the $n=3$
lobes which extends up to a triple point. A strong three-body attraction also affects the order of the MI-SF transition.
We have found first order transitions separating the $n=1$ and $n=3$ MI from the SF.
We have also analyzed the effect of finite temperature and found that transitions are still of first order at temperatures $T\sim J$.

\begin{acknowledgements}
We acknowledge fruitful discussions with H.R. Sadeghpour, M. Foss-Feig and A.M. Rey.
STR acknowledges the financial support of an NSF grant through ITAMP at Harvard University and the Smithsonian Astrophysical Observatory. JvS gratefully acknowledges the hospitality of ITAMP visitor program, where this work was initiated, and partial support from NSF and Tech-X Corporation.
\end{acknowledgements}

\end{document}